# Ultrafast exciton-polaron dynamics in 2D Ruddlesden–Popper lead halide perovskites

Anirban Mondal[1], Kwang Jin Lee[1,2,3]*, Seungmin Lee[4], Oui Jin Oh[4], Myeongsam Jen[5], Jun Hong Noh[4]*, Jong Min Lim[5]*, and Minhaeng Cho[1,6]*

[1] Center for Molecular Spectroscopy and Dynamics, Institute for Basic Science (IBS), Seoul 02841, Republic of Korea.
[2] Division of Semiconductor Physics, Korea University, Sejong 30019, Republic of Korea
[3] Department of Applied Physics, Graduate School, Korea University, Sejong 30019, Republic of Korea
[4] School of Civil, Environmental and Architectural Engineering, Korea University, Seoul 02841, Republic of Korea.
[5] Department of Chemistry and Green-Nano Materials Research Center, Kyungpook National University, Daegu 41566, Republic of Korea.
[6] Department of Chemistry, Korea University, Seoul 02841, Republic of Korea.
*e-mail: kwangjin1028@korea.ac.kr, jongmin@knu.ac.kr, mcho@korea.ac.kr

Two-dimensional Ruddlesden–Popper (2D RP) hybrid perovskites exhibit substantially higher chemical and structural stability than their three-dimensional (3D) counterparts, positioning them as promising candidates for next-generation optoelectronics. While quasiparticle dynamics in 3D perovskites are well studied, their 2D analogues remain comparatively underexplored. Here we systematically investigate the branching, dynamics, and interactions of free excitons (FEs) and exciton polarons (EPs) in $n = 1$ 2D RP perovskites using visible-range femtosecond transient absorption (TA) spectroscopy. We prepared $n = 1$ 2D RP perovskite thin films with varied organic spacers and distinct fabrication routes for comparative analysis. We find that the EP binding energy is 50–65 meV in $(BA)_2PbI_4$ and 37–39 meV in $(PEA)_2PbI_4$, consistent with spacer-layer–dependent coupling as corroborated by FTIR. We reveal a dynamic equilibrium between FEs and EPs that persists for tens of picoseconds. Notably, the TA signatures differ by fabrication route: films from the newly developed process show weaker Auger annihilation and a reduced hot-phonon bottleneck than those from the conventional route—trends consistent with fewer traps and impurities in the former. Coupled rate-equation modeling reproduces the transients and quantifies the processes of hot-carrier relaxation, exciton–exciton annihilation, exciton–phonon coupling, and FE↔EP interconversion. These results demonstrate that the chemical synthetic process (fabrication route) and spacer choice significantly influence EP stability and population balance, offering practical levers for engineering ultrafast photophysics in 2D perovskites and guiding the design of advanced optoelectronic devices.

**Introduction**

Since the discovery of graphene, two-dimensional (2D) semiconductors with a direct bandgap in the visible range, such as 2D transition metal dichalcogenides (TMDCs), have revolutionized applications in light-emitting and photovoltaic devices. These advancements are largely attributable to their tunable bandgap properties and pronounced exciton binding energies [1-4]. Furthermore, Ruddlesden–Popper (RP) type lead halide perovskites have emerged as promising multi-functional 2D semiconductors. They are garnering considerable research attention due to their tunable bandgap, suitability for solution-based processability, strong quantum confinement effects, and exceptional photoluminescence quantum yield.

2D RP perovskites are characterized by the chemical formula $L_2A_{n-1}M_nX_{3n+1}$ where L = ligand cation, M = metal cation, X = halide anion, and $n$ is an integer indicating the number of perovskite layers, consisting of lead halide octahedral sheets separated by insulating organic ligand cations [5, 6]. The insertion of bulky organic ligands into the perovskite lattice results in the formation of layered 2D perovskites with enhanced intrinsic stability [7-11]. Structural degradation can be significantly mitigated by sandwiching the inorganic octahedral layers between hydrophobic organic spacer layers [12]. As a result, 2D perovskites overcome the ligh, heat, and moisture instability issues that limit 3D counterparts in optoelectronic applications [13-17].

Apart from superior environmental stability, 2D RP perovskites also exhibit layer-tunable electronic properties, high exciton binding energies, slow hot-carrier relaxation, and strong nonlinear optical properties [18]. Their notable excitonic optical properties stem from quantum confinement and reduced dielectric screening from the 2D layered structure; electrons and holes in 2D perovskites inherently form strongly confined excitons. In addition, polaronic effects fundamentally influence exciton binding energy in these materials. It has been reported that the phonon coupling associated with polarons is in a unique intermediate regime where Fröhlich-like coupling dominates [19]. The presence of long-range interactions with the polar phonons and short-range lattice reorganization under photoexcitation results in exciton polarons (EPs) in 2D RP perovskites. While electron–phonon coupling is well-documented in quasi-2D and 3D lead halide perovskites [20-22], formally describing these quasiparticles remains challenging. Such descriptions are of fundamental importance for a complete understanding of excitonic behavior in 2D perovskites.

Subsequent studies have revealed exciton–phonon coupling in quasi-2D RP perovskites using cryogenic photoluminescence spectroscopy [23]. Furthermore, time-resolved microwave conductivity measurements indicated that more than 50% of the photoexcited species are free carriers rather than excitons in quasi-2D $(BA)_2(MA)_{n-1}Pb_nI_{3n+1}$ (n = 2) perovskites (BA = butylammonium, MA = methylammonium)[24]. As the number of layers increases, the relative population of free carriers to excitons increases, while the exciton binding energy decreases [25]. More recently, phonon-dressed excitons (EPs) in 2D RP perovskites have been identified as large Fröhlich polarons via time-resolved photoluminescence [26]. Exciton-phonon coupling in quasi-2D layered RP-type perovskites has been further studied using two-dimensional electronic spectroscopy and transient grating [27]. However, a quantitative understanding of many-body excitonic interactions, such as the branching of free excitons (FEs) into EPs and the dynamic equilibrium between these species in the ultimate 2D perovskite limit (n = 1), is still lacking.

This article explores quasiparticle dynamics in 2D RP lead iodide perovskites ($n = 1$): $(BA)_2PbI_4$ and $(PEA)_2PbI_4$ (BA = $C_4H_9NH_3^+$ and PEA = $C_6H_5C_2H_4NH_3^+$), which exhibit markedly different structural and photophysical properties owing to the distinct nature of their organic spacer cations. As shown in **Figures 1a and 1b,** the bulky phenethylammonium (PEA) cation in $(PEA)_2PbI_4$ increases interlayer separation and enhances structural rigidity, leading to strong dielectric confinement, higher exciton binding energy, sharp excitonic resonances, and enhanced photostability. In contrast, the shorter butylammonium (BA) cation in $(BA)_2PbI_4$ allows closer inorganic layer stacking and reduces lattice rigidity, thereby weakening dielectric confinement, broadening spectral linewidths, and relatively lowering exciton binding energy [19, 28, 29-32].

Furthermore, we conduct an in-depth comparison study using a recently developed route—selective iodoplumbate cold casting (SICC)—in which thin films are cast at ambient temperature and pressure from an acetonitrile (ACN): N-methyl-2-pyrrolidone (NMP) (5:4) solvent mixture, yielding kinetically stabilized 2D perovskite phases that display out-of-plane-favored orientation and improved vertical charge transport relative to N,N-dimethylformamide (DMF): dimethyl sulfoxide (DMSO) controls (see Methods). We directly compare $(BA)_2PbI_4$ and $(PEA)_2PbI_4$ prepared by SICC with films made by a conventional DMF:DMSO route. For brevity, samples are denoted $(BA)_2PbI_4$−S, $(BA)_2PbI_4$−C, $(PEA)_2PbI_4$−S, and $(PEA)_2PbI_4$−C (S = SICC; C = conventional). We then evaluate how the processing —solvent system and crystallization pathway—governs exciton–phonon coupling, polaron binding energy, and ultrafast carrier dynamics. We analytically determined the population ratio of EPs to FEs and extracted the polaron binding energies. Moreover, we numerically solved the coupled differential rate equations describing each photoexcited species and found excellent agreement between simulations and experimental observations. Our synthesis strategy and experimental findings highlight the potential of 2D RP perovskites as a feasible candidate for novel optoelectronic applications.

## Results

2D RP-type perovskites were synthesized using the precursor spin-coating method with a slight modification (see "Methods" for details), which is different from the literature [33, 34]. Precursor solutions were prepared in polar aprotic co-solvents and spin-coated into thin films; SICC yielded as-cast films at ambient conditions, while control films required thermal annealing. The inorganic octahedral layer is sandwiched between two types of spacer layers (BA and PEA) with different grain sizes tuned by various solvents.

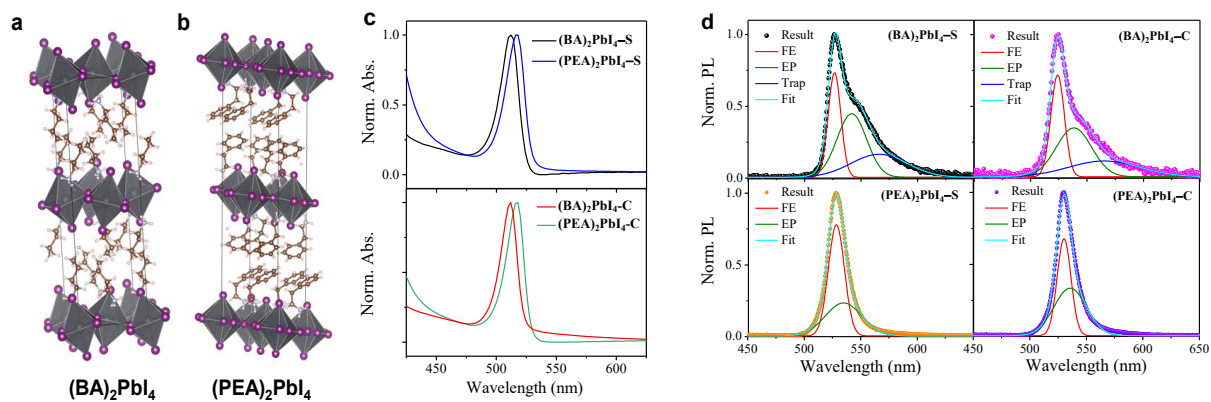

**Figure 1. Crystal structures of two 2D RP-type perovskites a,** $(BA)_2PbI_4$, **b,** $(PEA)_2PbI_4$. Optical absorption spectra (**c**) and Photoluminescence (PL) spectra (**d**) of four 2D RP perovskites $(BA)_2PbI_4$−S, $(BA)_2PbI_4$−C, $(PEA)_2PbI_4$−S, and $(PEA)_2PbI_4$−C, where S and C denote SICC and conventional DMF:DMSO methods, respectively. Gaussian functions convoluted PL spectra correspond to the free exciton (FE), exciton-polaron (EP), and trap states.

XRD measurements show that all thin film samples have a crystal structure with a sharp peak along the [002] direction at $2\theta = 6°$, signifying the existence of RP-type phase with $n = 1$ as shown in **Supplementary Figure 1**. The atomic force microscope (AFM) image reveals different grains of 2D perovskites, whose grain size depends on the fabrication method. Switching from the conventional DMF:DMSO route to the SICC (ACN:NMP) route yields smaller grains for both $(BA)_2PbI_4$ and $(PEA)_2PbI_4$ (**Supplementary Figure 2**).

**Figure 1c** shows the optical absorption spectra with a strong and sharp peak centered at 512 nm and 516 nm for $(BA)_2PbI_4$ and $(PEA)_2PbI_4$, respectively, attributing to the existence of numerous excitons over free carriers upon photoexcitation [35, 36]. The exciton binding energy is approximately 300 meV for $n = 1$ perovskites, as reported previously [37]. Our analysis notes a slight redshift in the optical bandgap due to the change in the organic spacer cation from BA to PEA in 2D perovskites.

**Figure 1d** presents the photoluminescence (PL) spectra of spacer-layer–dependent 2D perovskites, revealing distinct spectral signatures that provide insight into their quasiparticle dynamics. Newly fabricated samples $(BA)_2PbI_4$ –S and $(PEA)_2PbI_4$ –S exhibit higher PL intensity than $(PEA)_2PbI_4$ –C and $(PEA)_2PbI_4$ –C (**Supplementary Figure 3**), which also supports that the SICC method improves optical properties of 2D RP perovskites. In $(BA)_2PbI_4$ –S and $(BA)_2PbI_4$ –C, sharp emission peaks are observed at ~525 nm, while $(PEA)_2PbI_4$ –S and $(PEA)_2PbI_4$ –C exhibit peaks at ~529 nm, consistent with reported differences in grain sizes of the two systems [38, 39]. The emission energies correspond with the absorption spectra, indicating that $(BA)_2PbI_4$ has a slightly larger bandgap than $(PEA)_2PbI_4$. Importantly, the observed

PL redshift of ~60 meV is substantially larger than the typical 10–15 meV reported for 3D perovskites [40-42], which we attribute to the intrinsic polaronic nature of ionic lead–halide perovskites [43].

To quantify the spectral features, we fitted the PL profiles with sums of Gaussians. For $(BA)_2PbI_4$–S and $(BA)_2PbI_4$–C, three components were required: free exciton emission (FE, red line), exciton–polaron emission (EP, green line), and an additional red-shifted component attributed to trap-state emission (blue line). The overall fits are shown as cyan curves (Fit). The PL spectra of $(PEA)_2PbI_4$–S and $(PEA)_2PbI_4$–C are well described by two components, FE (red) and EP (green), with no significant contribution from trap states. The stronger EP component in $(BA)_2PbI_4$ relative to $(PEA)_2PbI_4$ indicates enhanced exciton–phonon coupling, while the presence of a distinct trap-related peak in $(BA)_2PbI_4$ but not in $(PEA)_2PbI_4$ highlights their contrasting trap densities. This difference arises from the reduced rigidity of BA-based perovskites, which enhances lattice disorder and fluctuations, facilitating the formation of shallow and deep traps that accelerate nonradiative recombination and shorten carrier lifetimes [19, 28, 29]. Conversely, the more rigid PEA-based structure suppresses trap formation, resulting in stable excitonic emission with fewer nonradiative losses. Taken together, we demonstrate how the choice of organic spacer—BA vs. PEA—dictates exciton–phonon coupling, polaron stabilization, and trap-state formation, ultimately governing the photophysical properties and optoelectronic performance of layered perovskites.

Complementary FTIR spectroscopy identifies a spacer-layer–dependent vibrational signature, with an additional band at 2957 cm$^{-1}$ attributed to an asymmetric C–H stretching mode in $(BA)_2PbI_4$, which is not observed in $(PEA)_2PbI_4$ (**Supplementary Figure 4**) [44, 45]. These measurements consistently show that $(BA)_2PbI_4$ has more prominent exciton–phonon interactions compared to $(PEA)_2PbI_4$, yielding relatively larger polaron binding energies [30-32].

To study quasi-particle interactions and dynamics, we performed transient absorption (TA) measurements at room temperature. Experimental details are provided in the Methods section. TA contour maps for all samples, recorded under 400 nm photoexcitation over varying pump fluences ($f$), are presented in the Supplementary Information (**Supplementary Figures 5-8**). Figure 2 displays the TA contour maps at a pump fluence of $f$ = 50 μJ cm$^{-2}$, revealing a pronounced ground state bleaching (GSB) at 516 nm for $(BA)_2PbI_4$ and 520 nm for $(PEA)_2PbI_4$, attributable to state filling of the FE resonance. A red-shifted photoinduced absorption (PIA) band appears near 530 nm. Notably, this PIA is already present at the lowest $f$ (20 μJ cm$^{-2}$; **Supplementary Figures 5-8**), which rules out a biexciton origin [46-49]. We can also exclude the optical Stark effect, as the derivative-like feature persists well beyond the pump–probe temporal overlap and the excitation is far from resonance. We, therefore, attribute the PIA to EPs, indicating that photoexcitation produces a mixture of FEs and EPs. Remarkably, EP signatures persist for tens of picoseconds at room temperature across all grain sizes and spacer layers.

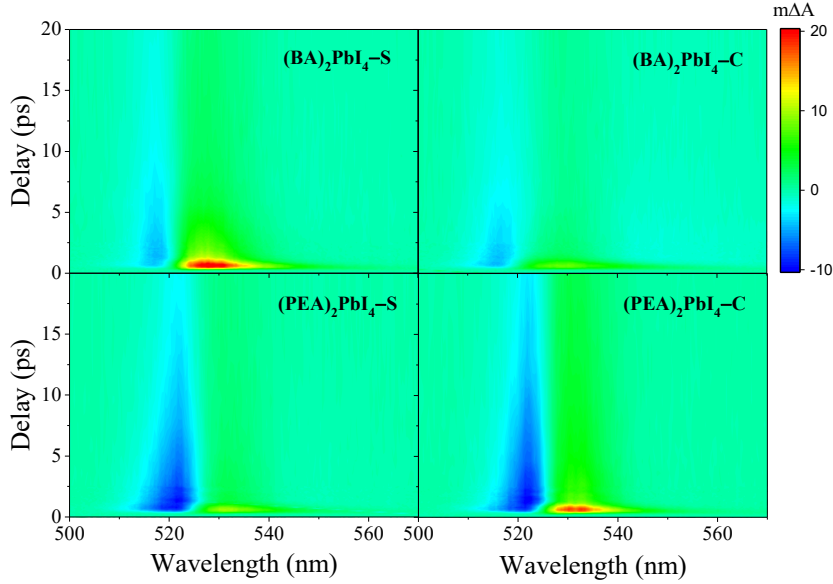

**Figure 2.** Contour plots TA data of $(BA)_2PbI_4$−S, $(BA)_2PbI_4$−C, $(PEA)_2PbI_4$−S, and $(PEA)_2PbI_4$−C with 400 nm photoexcitation at the pump fluence of 50 μJ/cm$^2$.

In **Figure 3**, we plot the spectral cross-sections at selected pump-probe time delays for $f$ =50 μJ cm$^{-2}$. From sub-ps to tens of ps, the TA spectra exhibit an asymmetric, derivative-like lineshape: a PIA band on the low-energy side of the band edge and a bleach on the high-energy side, reflecting an EP resonance [46, 49]. The FE and EP populations rise promptly from hot carriers and reach maxima near 0.8 ps. Consistent with PL analysis indicating a higher trap-state density in $(BA)_2PbI_4$, a broad negative feature appears around 550 nm for $(BA)_2PbI_4$−C (**Fig. 3b**), which we attribute to stimulated emission (SE) from trapped polarons. Interestingly, we find that the SE feature disappears in $(BA)_2PbI_4$−S (**Fig. 3a**). This is understandable since newly synthesized $(BA)_2PbI_4$−S possesses comparatively lower trap density.

In $(PEA)_2PbI_4$, where the trap density is negligible, no such SE is observed (Fig. 3c and 3d). The kinetics at 550 nm exhibit a rapid positive excursion followed by a negative signal, reflecting exciton-polaron trapping followed by trap-assisted SE. In $(BA)_2PbI_4$−S, the lower trap density results in weaker trap-mediated SE; nevertheless, the PIA amplitude is larger, reflecting a higher EP population, consistent with PL fitting (**Supplementary Figure 9**).

To examine the dynamics of FEs and EPs separately, we assess their using a TA spectrum model that sums two overlapping Gaussians[49]. At 0.8 ps, the TA spectrum is well described by [46]:

$$\Delta A(x) = A_p \exp\left[\left(\frac{x - x_c - E_b}{w_p}\right)^2\right] - A_e \exp\left[\left(\frac{x - x_c}{w_e}\right)^2\right] \qquad (1)$$

where $x$ is the spectral axis (photon energy), $A_e$ and $A_p$ are the amplitudes of GSB and PIA, respectively, $x_c$ is the FE resonance, $E_b$ is the EP binding energy (red-shift relative to $x_c$), and $w_e$ and $w_p$ are corresponding widths. We define the branching coefficient $\beta$ as the ratio of EP to FE populations, estimated from the integrated areas of the two Gaussians:

$$\beta \equiv \frac{N_{pol}}{N_{ex}} \approx \frac{A_{PIA}}{|A_{GSB}|} \qquad (2)$$

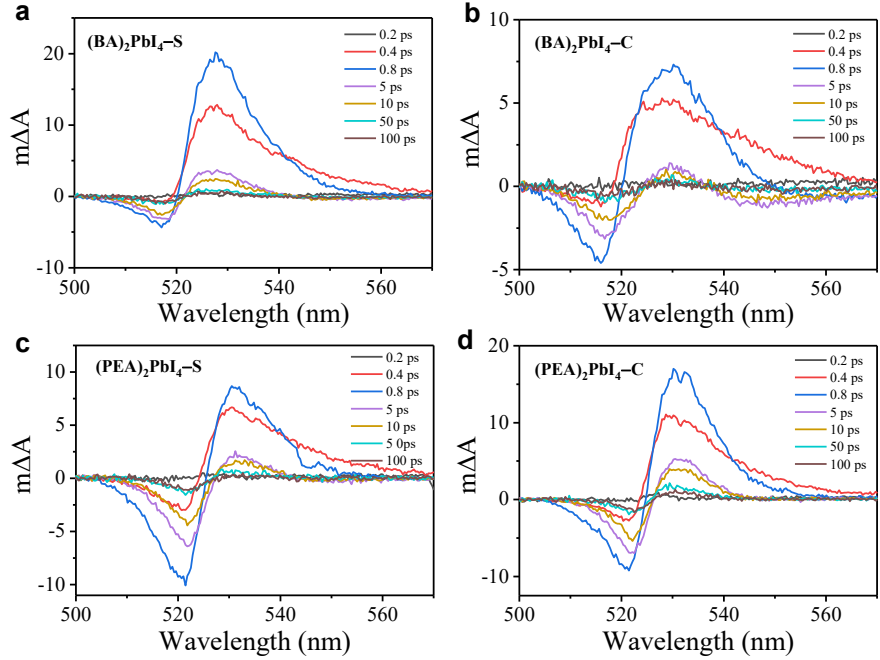

**Figure 3. Delay time-dependent spectral features of four 2D perovskites acquired from the spectral cross-section of transient absorption (TA) contour map at the pump fluence of 50 μJ/cm$^2$**: (a-d) Spectra at 0.2 ps to 100 ps are given for all the perovskites (perovskite names are mentioned in the left upper corner of each image.

In **Figure 4**, the polaron-induced absorption (PIA, dark green) and the free exciton bleach (GSB, blue) are displayed; the branching coefficient $\beta$ is obtained by integrating these components. The extracted polaron binding energies $E_b$ from Eq.(1) are 50 – 65 meV for (BA)$_2$PbI$_4$ and 37−39 meV for (PEA)$_2$PbI$_4$. These results agree with previous reports: (PEA)$_2$PbI$_4$ exhibits a Fröhlich constant α of ~1.5–2.0 with a binding energy of ~20–40 meV, whereas (BA)$_2$PbI$_4$ shows stronger coupling (α ~2.0–2.5) and larger binding energies (~40–70 meV) [29-32]. Notably, all $E_b$ values exceed room temperature energy (~25 meV) and are larger than those typically reported for 2D TMDCs, despite the latter's large exciton binding energies [50]. Using Eq.(2), we obtain $\beta$ = 2.41 and 2.22 for (BA)$_2$PbI$_4$−S and (BA)$_2$PbI$_4$−C, respectively; for (PEA)$_2$PbI$_4$−S and (PEA)$_2$PbI$_4$−C, $\beta$=1.17 and 1.56, respectively (**Supplementary Table 1**). These indicate that (BA)$_2$PbI$_4$ sustains a substantially higher EP fraction than (PEA)$_2$PbI$_4$.

The two-Gaussian decomposition informs the selection of probe energies for selective monitoring of individual quasiparticle channels. Probing at 2.33 eV (532 nm) and 2.42 eV (514 nm) allows exclusive monitoring of EP and FE dynamics, respectively. Accordingly, we present *f*-dependent dynamics at a probe wavelength of 514 nm for the FE and 532 nm for the EP (**Supplementary Figure 10**). Overall, the TA time profiles are well described by a tri-exponential model across all *f* s (**Supplementary Tables 2–9**). We tentatively assign the three decay time components as follows: the shortest time ($\tau_1$ < 1 ps) reflects hot carrier relaxation and transition from FE to EP conversion; the intermediate one ($\tau_2$<100 ps) arises from

exciton–exciton annihilation and dissociation of EP to FE; and the longest one ($\tau_3$>200 ps) corresponds to radiative exciton recombination.

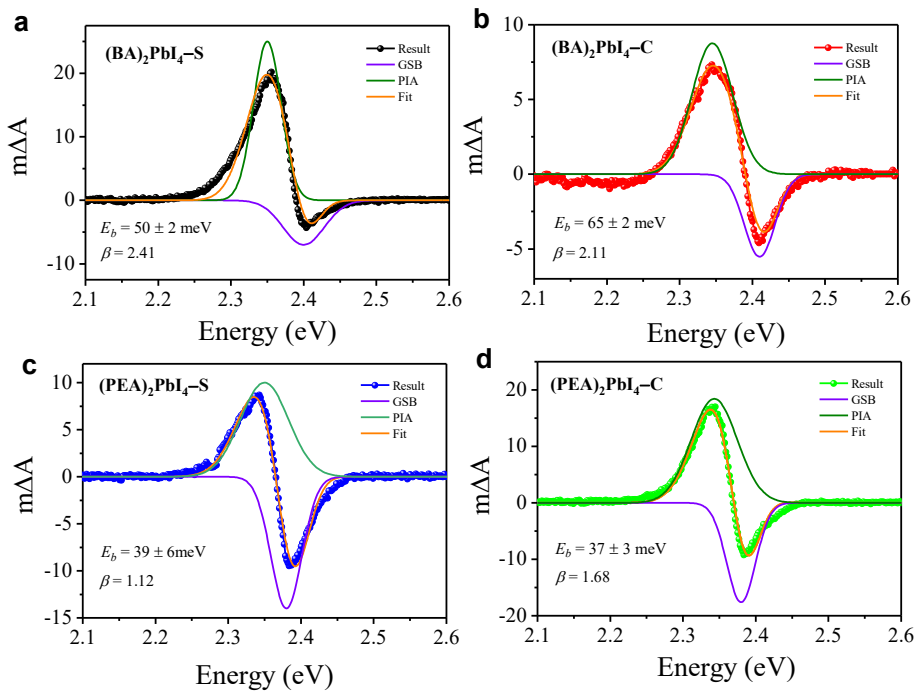

**Figure 4. Extraction of exciton–polaron binding energy from TA spectra at 0.8 ps.** (a–d) Experimental spectra (green symbols) fitted with a two-Gaussian decomposition: exciton–polaron photoinduced absorption (PIA, dark green) and free exciton ground-state bleach (GSB, blue). The resulting composite fit (red) reproduces the data. Pump fluence: 50 μJ cm$^{-2}$ at 400 nm. For all samples, the area ratio $\beta$ (i.e., exciton-polaron to free exciton fraction) exceeds unity. The extracted polaron binding energy $E_b$ for each sample is reported in the upper-left of each panel.

The fitted time constants $\tau_1$ and $\tau_2$ corresponding to FE dynamics as a function of $f$ are plotted in **Figure 5a** at 514 nm probe. For (BA)$_2$PbI$_4$–S, the free exciton bleach at the lowest $f$ is well captured by a single-exponential decay. With increasing $f$, additional time components are required to fit the results, and its TA dynamics curves are well fitted by a tri-exponential function (**Supplementary Tables 2–5**) when $f$ =50 μJ/cm$^2$. For (BA)$_2$PbI$_4$–C, tri-exponential decays are observed at all $f$s. A decrease in $\tau_2$ with increasing $f$ is a signature of exciton–exciton (Auger) annihilation. The increase in $\tau_1$ with $f$ is likely due to the hot-phonon bottleneck effect. The TA dynamics curves of (PEA)$_2$PbI$_4$-S and (PEA)$_2$PbI$_4$-C exhibit tri-exponential decay at all $f$. In both cases, $\tau_2$ decreases with $f$ indicating active exciton–exciton annihilation. Similar to (BA)$_2$PbI$_4$–C, (PEA)$_2$PbI$_4$–C also shows an increase in $\tau_1$ with $f$. The weaker Auger effect and phonon bottleneck in (BA)$_2$PbI$_4$–S and (PEA)$_2$PbI$_4$–S compared to (BA)$_2$PbI$_4$–C and (PEA)$_2$PbI$_4$–C indirectly support that newly synthesized samples are relatively freer from defects and impurities.

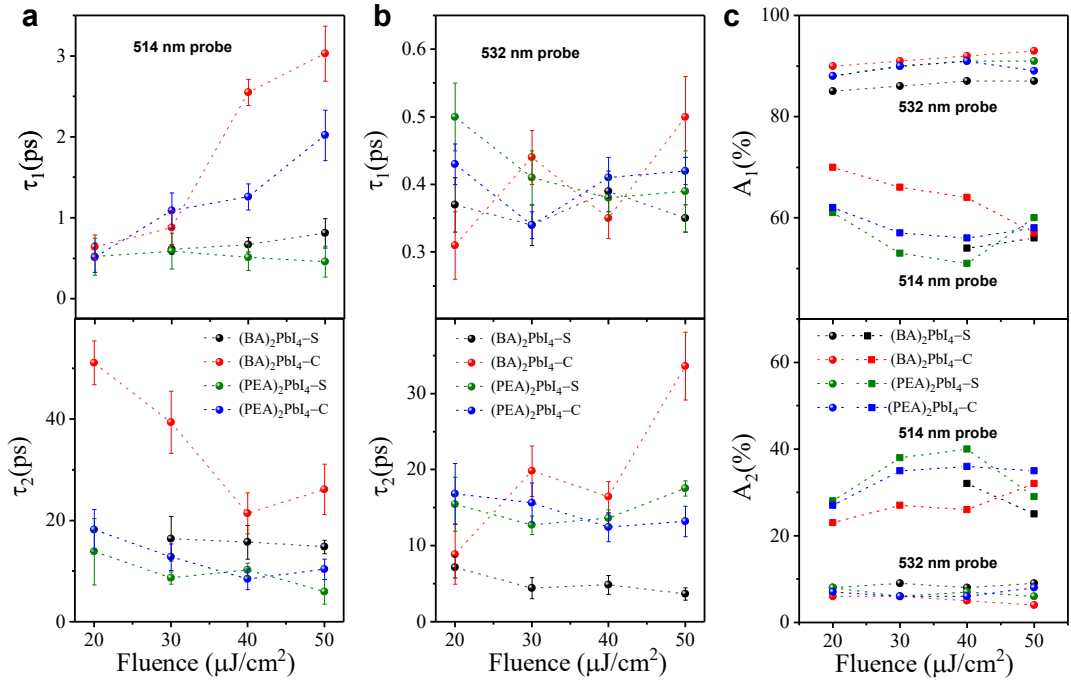

**Figure 5. a, b** Plots of the decay time constants ($\tau_1$ and $\tau_2$) for the multi-exponential fit to the free exciton (514 nm probe **a**) and exciton polaron (532 nm probe **b**) kinetic profiles and their amplitudes ($A_1$ and $A_2$) as a function of pump fluence **c**.

In **Figure 5b**, we also plot the fitted time constants $\tau_1$ and $\tau_2$ as a function of $f$ at 532 nm probe for analyzing EP dynamics. All samples exhibit tri-exponential decays (**Supplementary Tables 6–9**). In contrast to the FE dynamics, $\tau_1$ is essentially $f$-independent, supporting its assignment to hot-carrier relaxation that feeds the EP population. Except for $(BA)_2PbI_4-C$, no systematic $f$ dependency is observed for $\tau_2$, arguing against polaron–polaron annihilation; we therefore assign $\tau_2$ to dissociation of EPs to FEs. Significant increase in $\tau_2$ for $(BA)_2PbI_4-C$ might be due to relatively high impurity and defect levels. As shown in **Figure 5c**, interestingly, the amplitude of $\tau_1$ ($A_1$) is almost 90% for EP decay (532 nm probe), whereas it is around 50~70% for FE decay (514 nm probe). This implies that hot carriers undergo relaxation efficiently into stable EP states, while concurrent EP→FE back-conversion prevents EP accumulation; accordingly, EP–EP annihilation can be suppressed, consistent with the weak fluence dependence and smaller $A_2$ in the EP kinetics (532 nm probe) (**Supplementary Tables 2–9**). The slowest component is not analyzed further due to the finite temporal window.

Although multi-exponential fitting is widely used to obtain kinetic parameters from TA decays, it alone cannot conclusively determine dynamics because TA signals reflect several overlapping photophysical processes. For example, the PIA observed at 532 nm in all samples contains contributions from both hot-carrier and polaron-induced absorption; therefore, $\tau_1$ cannot be uniquely identified from multi-exponential fitting alone.

To undertake a comprehensive spectral and temporal analysis of the TA data, we begin by validating the underlying kinetic model. Guided by an energy-level (band) diagram, we cconstruct coupled rate equations for each species and solve them numerically to assess whether the simulated rate constants reproduce the experimentally extracted ones across all 2D perovskites. The model should also account for the experimentally observed many-body features, which we verify below.

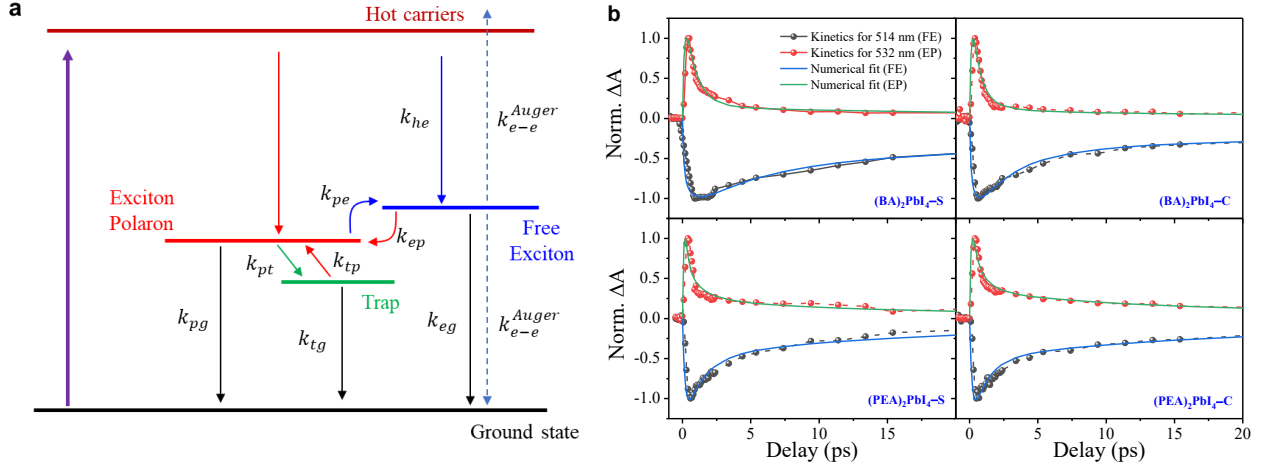

**Figure 6. a**, Energy-level diagram illustrating the coupled populations of free excitons and exciton polarons. Above-gap excitation generates hot carriers that relax to free excitons at rate $k_{he}$ exciton polarons at $k_{hp}$. The two species interconvert via $k_{ep}$(exciton → polaron) and $k_{pe}$ (polaron → exciton). Polarons can be trapped into sub-gap states at $k_{pt}$ and thermally released at $k_{tp}$; trapped polarons decay to the ground state at $k_{tg}$. Ground-state decay from free excitons and polarons occurs at $k_{eg}$ and $k_{pg}$, respectively. A bimolecular exciton–exciton Auger channel ($k_{e-e}^{Auger}$; dashed upward arrow) promotes a carrier to the hot-carrier manifold. **b,** TA kinetics (symbols) and numerical fits (solid lines) at representative probe wavelengths for all samples.

In the band diagram scheme shown in **Fig. 6a**, above-gap photoexcitation promptly generates hot carriers (violet upward arrow). **Figure 6a** summarizes the kinetic scheme used to model the TA dynamics. A femtosecond above-gap pump (violet upward arrow) creates hot carriers, which relax through two parallel channels: formation of free excitons at rate $k_{he}$ (blue downward arrow) and direct formation of exciton-polarons at $k_{hp}$ (red downward arrow). The two bound species remain coupled by bidirectional interconversion, from free exciton to polaron with $k_{ep}$ and vice versa with $k_{ep}$ (curved arrows). A sub-gap trap manifold (green) captures polarons with $k_{tp}$ and can thermally release them with $k_{pt}$; trapped populations may also decay to the ground state with $k_{tg}$. Return to the ground state occurs from both bound states: $k_{eg}$ for free excitons and $k_{pg}$ for exciton-polarons (black downward arrows). At elevated densities, a bimolecular exciton–exciton (Auger) annihilation channel depletes the free exciton population (dashed pathway). The dashed upward arrow denotes exciton–exciton (Auger) annihilation with effective rate $k_{e-e}^{Auger}$, whereby the recombination energy of one exciton promotes another carrier into the hot-carrier manifold; these carriers then cool back to excitons or polarons via $k_{he}$ and $k_{hp}$. This network of processes maps onto the observed TA signatures.

The kinetics are obtained by numerically solving four coupled first-order differential equations for the populations of each energy state with respect to time delay $t$ given by

$$\frac{dP_h}{dt} = -(k_{he} + k_{hp})P_h + \frac{1}{2}k_{e-e}^{Auger}P_e^2 \qquad (3a)$$

$$\frac{dP_e}{dt} = k_{he}P_h + k_{pe}P_p - (k_{ep} + k_{eg})P_e - k_{e-e}^{Auger}P_e^2 \tag{3b}$$

$$\frac{dP_p}{dt} = k_{hp}P_h + k_{ep}P_e + k_{tp}P_t - (k_{pt} + k_{pe} + k_{pg})P_p \tag{3c}$$

$$\frac{dP_t}{dt} = k_{pt}P_p - k_{tp}P_t - k_{tg}P_t \tag{3d}$$

where $P_h, P_e, P_p,$ and $P_t$ denote the populations of hot carriers, free exciton, exciton polaron, and trapped exciton polaron, respectively. We also note the relation given by $k_{pe} = k_{ep}\exp[-\Delta E/k_B T]$, where $\Delta E$ is the polaron binding energy. Equations (3a)-(3d) describe, respectively, the generation/relaxation of hot carriers and the formation/decay of FEs, EPs, and trapped polarons. Initial conditions right after the pump (at $t = 0$) are given by $P_f(0) = 1, P_e(0) = 0, P_p(0) = 0,$ and $P_t(0) = 0$ (assuming that the populations are normalized). Solving the rate equations with these initial conditions yields $P_h(t), P_e(t),$ and $P_p(t)$.

**Table 1.** Fitted kinetic parameters from numerical modelling and simulation ($f = 50\ \mu J/cm^2$)

| Sample | $k_{he}$ (ps$^{-1}$) | $k_{hp}$ (ps$^{-1}$) | $k_{e-e}^{Auger}$ (ps$^{-1}$) | $k_{ep}$ (ps$^{-1}$) | $k_{pe}$ (ps$^{-1}$) | $k_{pt}$ (ps$^{-1}$) |
|---|---|---|---|---|---|---|
| (BA)$_2$PbI$_4$−S | 3.05 | 4.46 | 0.22 | 0.59 | 0.25 | 0.71 |
| (BA)$_2$PbI$_4$−C | 3.34 | 5.23 | 0.13 | 0.84 | 0.31 | 0.89 |
| (PEA)$_2$PbI$_4$−S | 2.85 | 5.78 | 0.63 | 2.62 | 0.95 | 0.13 |
| (PEA)$_2$PbI$_4$−C | 3.12 | 5.34 | 0.56 | 1.86 | 0.80 | 0.15 |

However, the TA signal does not directly report these populations; rather, it reflects a weighted superposition of contributions from the participating states. We thus make connections between these populations with time-resolved TA signals. We suppose that the TA signal ($\Delta A$) is given as the product of a weighting factor $a$ and the population $P$ of a given state. For example, within a linear-response model at a fixed probe energy, the TA signal can be represented by a linear combination of each population:

$$\Delta A(t) = a_h P_h(t) + a_e P_e(t) + a_p P_p(t) \tag{4}$$

where the weighting factors $a_h, a_e,$ and $a_p$ (probe-wavelength dependent) are the sum of the GSB and PIA channels of each state.

Fitting Eqs. (3a) – (3d) together with these weighting factors in Eq. (4) reproduces, simultaneously, the exciton bleach (514 nm) and polaron-induced absorption (532 nm) transients for all samples (**Fig. 6b**). The simulations reproduce the concurrent formation of FEs and EPs within ~ 0.3 ps and capture the measured transients and provide quantitative estimates of PIA contributions to the TA signals of each species (**Table 1**). The significantly larger $k_{pt}$ extracted for (BA)$_2$PbI$_4$ relative to (PEA)$_2$PbI$_4$ indicates more active trap-mediated pathways, consistent with a higher trap-state density in the BA films. The analysis indicates that hot-carrier contributions to the PIA are approximately 7–11% at 514 nm and 4–6% at 532 nm (**Supplementary Table 10**).

**Discussion**

We fabricated uniform 2D RP perovskite thin films via SICC and by a conventional DMF:DMSO route and mapped their quasiparticle dynamics with femtosecond TA spectroscopy. Our combined experiments and modeling establish a coherent photophysical picture for 2D RP-type perovskites. Strong exciton–phonon coupling gives rise to EPs that persist for tens of picoseconds at room temperature. Two-Gaussian analysis yields $E_b$ of 50 – 65 meV for (BA)$_2$PbI$_4$ and 37−39 meV for (PEA)$_2$PbI$_4$, consistent with a BA-specific 2957 cm$^{-1}$ FTIR mode indicating stronger exciton–phonon coupling. Fluence-dependent TA traces provide further evidence of EPs in (BA)$_2$PbI$_4$ relative to (PEA)$_2$PbI$_4$. Observations of a weaker Auger effect and a reduced phonon bottleneck in samples fabricated by SICC, compared to those prepared by the conventional route, are consistent with the fact that the SICC method yields fewer defects and impurities. To quantitatively rationalize these observations, we constructed an energy-level–guided kinetic model and solved coupled rate equations for hot carriers, FEs, EPs, and trapped states. The numerical simulations capture the measured transients and provide quantitative estimates of GSB and PIA contributions to the TA signals of each species. We note that a coupled rate-equation model reproduces concurrent FE/EP formation within ~0.3 ps and quantifies PIA contributions (~7–11% at 514 nm; 4–6% at 532 nm). Overall, the chemistry processing route and spacer layer jointly control the EP fraction, $E_b$, and trap-mediated lifetimes (>100 ps), offering practical levers for optimizing 2D perovskite optoelectronics. Although demonstrated here for 2D RP perovskites, this framework is readily extendable to DJ-type perovskites and other quantum materials, offering a general route to disentangle photophysical pathways and to inform the design of stable, on-demand optoelectronic devices. Moreover, engineering the electromagnetic environment by introducing nanocomposite structures offers a promising handle for controlling EP dynamics, as shown in related hybrid systems [51-55].

**Acknowledgments**
This research was supported by the National Research Foundation of Korea (NRF) grants funded by the Korean government (No. RS-2025-00558173 and No. RS-2023-00301914). This work was also supported by a Korea University Grant. This work was supported by the Korea Institute of Energy Technology Evaluation and Planning (KETEP) grant funded by the Ministry of Trade, Industry, and Energy (20214000000680).

**Additional Information**

**Competing Financial Interests**
The authors declare no competing financial interests.

**Author Contributions**
K.J.L. and M.C. designed the project and coordinated the overall experiments. S.L. and O.J.O. prepared 2D perovskite samples and performed XRD and AFM measurements supervised by J.H.N. A.M. carried out basic optical measurements as well as ultrafast experiments with the help of J.M.L and the supervision of M.C.. A.M., K.J.L., and J.M.L. systematically analyzed the overall data with the help of M.C. A.M., K.J.L., and M.C. wrote the manuscript. All authors commented on the paper.

## Methods
### Chemicals
Butylammonium iodide (BAI) and phenethylammonium iodide (PEAI) were purchased from Great Solar and used as received. Lead iodide ($PbI_2$) was purchased from Tokyo Chemical Industry Co., Ltd. (TCI) and used as received. Acetonitrile (ACN), N-methyl-2-pyrrolidone (NMP), N,N-dimethylformamide (DMF), and dimethyl sulfoxide (DMSO) were obtained as anhydrous grades from Sigma-Aldrich and used as received unless otherwise noted.

### Synthesis
### Control film preparation
Stoichiometric n = 1 precursor solutions (0.2 M) were prepared by dissolving the spacer iodide (BAI or PEAI) and $PbI_2$ in DMF: DMSO at an 8:1 (v/v) ratio, filtered, and spin-coated onto pre-cleaned ITO-coated glass at 5000 rpm for 30 s. Diethyl ether (1 mL) was rapidly dispensed onto the spinning substrate 10 s before the end of the program. After deposition, films were thermally annealed at 100 °C for 5 min.

### SICC film preparation
Stoichiometric n = 1 precursor solutions (0.2 M) were prepared by dissolving the spacer iodide (BAI or PEAI) and $PbI_2$ in an ACN:NMP co-solvent at a 5:4 (v/v) ratio, filtered, and spin-coated onto pre-cleaned ITO-coated glass at 5000 rpm for 30 s. Diethyl ether (1 mL) was rapidly dispensed onto the spinning substrate 10 s before the end of the program to trigger crystallization, and films were formed at ambient temperature and pressure without post-annealing unless otherwise noted.

### Optical absorption and Photoluminescence spectroscopy
UV-Vis electronic absorption spectra for thin film samples were measured using Duetta (Horiba Scientific) at room temperature. The xenon arc lamp has been used as an excitation source. Emission spectra of 2D perovskite thin films were recorded using a Duetta (Horiba Scientific). Our samples were excited at 400 nm, and data were collected at ambient temperature. The Silicon photodiode in the detector channel can detect wavelengths from 250 to 1100 nm.

### X-ray diffraction
XRD measurements were performed on a Rigaku SmartLab diffractometer equipped with a Cu Kα source ($\lambda$ = 1.5406 Å). Scans were collected in Bragg–Brentano reflection geometry with a step size of 0.02° and a scan rate of 1° $min^{-1}$; unless otherwise noted, the 2θ range was selected to fully capture the (00l) series and impurity signatures characteristic of n = 1 RP phases. Thin-film specimens on an ITO substrate were measured at room temperature under ambient conditions without sample spinning.

### Atomic force microscopy
Atomic force microscopy (AFM) topography was collected on a Park Systems NX10 housed in an inert-atmosphere glovebox for 2D perovskite films deposited on ITO-coated glass at room temperature, using true non-contact mode to minimize tip–sample interaction artifacts. Images were acquired over 5 μm × 5 μm scan areas.

### Transient absorption (TA) pump-probe spectrometer
The ultrafast dynamics measurements were investigated using a custom-built transient absorption (TA) spectrometer operated in transmission geometry. A Yb:KGW laser (PHAROS) produced 100 fs pulses at 1030 nm with a repetition rate of 50 kHz and an average output power of 6 W. The laser output was split into two beams. One portion was focused into a YAG crystal to generate a white-light continuum, which served as the probe. The other portion was directed into an optical parametric amplifier (Orpheus) to produce 800 nm pulses. After passing through a mechanical chopper, this fundamental beam was frequency-doubled in a BBO crystal to generate 400 nm pulses, which were used as the pump. Pump and probe beams were spatially overlapped on the sample, with the pump beam diameter set to ~100 μm. The transmitted probe light was collected via an optical fiber and analyzed with a spectrometer (OCEAN FX). All experiments were performed at room temperature under ambient conditions.

**Fourier transform infrared spectroscopy**

Infrared spectra were collected with a PerkinElmer Frontier FT-IR Spectrometer in transmission mode. The spectrometer contains mid-IR source lamp (30-8000 cm$^{-1}$) and LiTaO$_3$ detector (370-15700 cm$^{-1}$). Samples on a quartz substrate were attached to the IR sample cell for measurement. In each experiment, four scans were performed with 4 cm$^{-1}$ spectral resolution.